\documentclass[fleqn,usenatbib]{mnras}

\usepackage{newtxtext,newtxmath}

\usepackage[T1]{fontenc}
\usepackage{ae,aecompl}

\usepackage{graphicx}	
\usepackage{epstopdf}
\usepackage{amsmath}	
\usepackage{amssymb}	

\title[Characterizing ulGRBs]{Can we quickly flag Ultra-long Gamma-Ray Bursts?}

\author[B. Gendre et al.]{
B. Gendre,$^{1,2,3,4}$\thanks{E-mail: bruce.gendre@uwa.edu.au}
Q.T. Joyce,$^{1}$
N.B. Orange,$^{2,5}$
G. Stratta,$^{6,7}$
J.L. Atteia,$^{8}$
and M. Bo\"er$^{9}$
\\
$^{1}$University of the Virgin Islands, College of Science and Mathematics, 2 John Brewers Bay, 00802 St Thomas, V.I., USA\\
$^{2}$Etelman Observatory, St Thomas, V.I., USA\\
$^{3}$OzGrav-UWA, University of Western Australia, School of Physics, M013, 35 Stirling Highway, Crawley, WA 6009, Australia\\
$^{4}$Present address: University of Western Australia\\
$^{5}$OrangeWave Innovative Science, LLC, Moncks Corner, SC 29461, USA\\
$^{6}$INAF-Osservatorio di astrofisica e scienza dello spazio, via P. Gobetti 93/3, 40129, Bologna, Italy\\
$^{7}$INFN, Sezione di Firenze, I-50019 Sesto Fiorentino, Firenze, Italy\\
$^{8}$IRAP, Universit\'e de Toulouse, CNRS, CNES, UPS, 14 Avenue Edouard Belin, F-31400 Toulouse, France\\
$^{9}$ARTEMIS UMR 7250 UCA CNRS OCA, boulevard de l'Observatoire, CS 34229, 06304 Nice Cedex 04, France
}

\date{Accepted XXX. Received YYY; in original form ZZZ}

\pubyear{2018}

\begin{document}
\label{firstpage}
\pagerange{\pageref{firstpage}--\pageref{lastpage}}
\maketitle

\begin{abstract}
Ultra-long Gamma-Ray Bursts  are a class of high energy transients lasting several hours. Their exact nature is still elusive, and several models have been proposed to explain them. Because of the limited coverage of wide field gamma-ray detectors, the study of their prompt phase with sensitive narrow-field X-ray instruments could help in understanding the origin of ultra-long GRBs. However, the observers face a true problem in rapidly activating follow-up observations, due to the challenging identification of an ultra-long GRB before the end of the prompt phase. We present here a comparison of the prompt properties available after a few tens of minutes of a sample of ultra-long GRBs and normal long GRBs, looking for prior indicators of the long duration. We find that there is no such clear prior indicator of the duration of the burst. We also found that statistically, a burst lasting at least 10 and 20 minutes has respectively $28\%$ and $50\%$ probability to be an ultralong event. These findings point towards a common central engine for normal long and ultra-long GRBs, with the collapsar model privileged.
\end{abstract}

\begin{keywords}
gamma-ray burst: general -- methods: observational
\end{keywords}



\section{Introduction}
\label{intro}

From an etymological point of view, Gamma-Ray Bursts (GRBs) are observationally defined events: they are bursts of $\gamma$-ray photons. The phenomenon was discovered in the late 1960's by the Vela satellites \citep{kle73}, and since then has been studied with passion by generations of high-energy astronomers with various instruments dedicated to GRB research \citep[e. g. KONUS, PHEBUS, BATSE, BeppoSAX, HETE-2, Swift, ][]{maz81, bar92, fis94, pir98, ric03, geh04}. We now know that in fact GRBs encompass different kinds of physical events. It is understood that GRBs \citep[see][for a review]{mes06} are fantastic explosions at cosmological distances due to either the merging of two compact objects \citep{eic89}, or the death of supermassive stars \citep{woo93}. Soft gamma-ray repeaters \citep{maz82}, or tidal disruption events \citep[such as Swift J164449.3+573451][]{bur11}, are also visible in gamma-rays and could be wrongly mistaken for GRBs.

A few years ago, a new ultra-long class of GRBs has been identified \citep{gen13}, and several authors have proposed various bursts to be classified as such events \citep[e.g.][]{lev14, cuc15, lie16}. The ultra-long GRB class is still mysterious, and at the moment they are only classifiable via observational properties \citep[see for instance][]{lev15, iok16}. These events can be explained with three main classes of progenitors: an ultra-massive stellar progenitor, very similar to Pop III stars \citep{suw11, nag12, mac13}; the tidal disruption of a dwarf star \citep{mac14}; and a newborn magnetar \citep{gre15}.

Disentangling among the progenitor models of ultra-long GRBs is difficult, as by construction they predict the same kind and level of emission \citep[see the interesting discussion in][for instance]{iok16}. To overcome such limitations, we are lacking key observations that could narrow the proposed models. \citet{str13} has already shown that the afterglow of ultra-long GRB 111209A is not different from any long burst one. Albeit of different types, ultra-long GRBs and normal long GRBs are associated with supernovae \citep{gre15}. The key to solving the nature of ultra-long GRBs, if it exists, likely resides in prompt phase multiwavelength observations. Most ultra-long GRBs last for a couple of hours \citep[see][]{boe15}. Though observing a phenomenon that last so long would seem relatively simple in this era of large-scale facilities, such data remains elusive because observational prior indicators that foreshadow these bursts are non-existent: most of these facilities being oversubscribed, each observational minute is extremely valuable, and obtaining unanticipated short term observations (in less than 5 hours) is highly competitive. A predictive method is thus necessary for observers to assess the probability that a given long GRB is in fact an ulGRB in order to perform radio, optical, or X-ray follow-up as early as possible during the prompt phase. We have studied the information available at the time of the trigger to see if such a method could be defined or if it was impossible. The purpose of this article is to report our findings.

We present in Sec. \ref{sample} the samples used for this study, which are classified as a: gold sample, made up of genuine ultra-long GRBs (ulGRBs hereafter); a silver sample, comprised of doubtful events; and a control sample of long GRBs. We then explore their spectral (Sec. \ref{spectral}) and temporal (Sec. \ref{timing}) properties, which are discussed in Sec. \ref{discu} before our conclusions. In the remainder of our manuscript, all measurement errors are given at the 90 \% confidence level, while statistical results are at the 3$\sigma$ level. A flat $\Lambda$CDM cosmological model (H$_0$ = 72, $\Lambda$ = 0.73) is used when needed.

\section{Sample selection}
\label{sample}

For this study, we focus on Swift data to build an homogeneous sample to avoid selection biases between different detectors. Observations available up to November 2017 were used to build our gold, silver, and control samples, based on event durations (see below).

As discussed in \citet{zha14} and \citet{boe15}, the separation between ulGRBs and long GRBs is not well defined, and various authors are using an ad-hoc separation value, complicating the comparison of the samples. \citet{lie16} for instance use emission lasting more than 1000 seconds in the $15-350$ keV energy range. \citet{boe15} used also a duration of 1000 seconds but in the $0.2-10$ keV band. Thus, the construction of our ulGRB (gold), and long GRB (control) samples were done carefully in order to avoid the "gray area" where both bursts could be present. We used this gray area to build a (silver) sample of events that could or could not be ulGRBs, as a blind sample for classification tests.

We measured event duration using the method of \citet{boe15}, which defines the prompt emission end point as the start of the steep decay phase (T$_x$), and thus the total duration of the event. While this is not the standard way (i.e., opposed to the usual T$_{90}$ measurement in $\gamma$-ray), it allows the samples to be unbiased toward the limited sensitivity of the BAT instrument in some cases.

As noted above, \citet{boe15} used the limit of T$_x = 10^3$s to discriminate long and ultra-long events. However, that value is not above the $3\sigma$ level of the distribution of long GRBs, and as such some these events could still be present. We considered T$_x = 10^3$s to be the lower limit of our silver sample. The value T$_x = 5 \times 10^3$s (where the $3\sigma$ limit is reached) was used to set the boundary between our silver and gold samples. For our control sample, we defined it as GRBs with a $T_{90}$ duration of at least 500s not belonging to any of the previous two groups to ensure that trivial bursts (i.e. those lasting a few seconds) are not used to test for ultra-long durations. These criteria were used to construct the gold, silver, and control samples presented in Tables \ref{table_gold}, \ref{table_silver}, and \ref{table_control}, respectively.

\begin{table}
	\centering
	\caption{The gold sample of ulGRBs with T$_x > 5 \times 10^3$s. We list selected properties.}
	\label{table_gold}
	\begin{tabular}{clll}
		\hline
		Name & Duration     & Duration    & Redshift\\
		     & (T$_{90}$,s) & (T$_{X}$,s) &  \\
		\hline
		GRB 101225A & > 7,000 &  5296  &	0.847\\
		GRB 111209A & 25,000  & 25,400 & 0.677\\
		GRB 121027A & > 6,000 &  8000  & 1.77\\
		GRB 130925A & 4,500   & 10000  & 0.35\\
		GRB 170714A &  420    & 16,600 & 0.793\\
		\hline
	\end{tabular}
\end{table}

\begin{table}
	\centering
	\caption{The silver sample of possible ulGRBs, selected with $5 \times 10^3$s $>$ T$_x > 10^3$s. We list selected properties.}
	\label{table_silver}
	\begin{tabular}{clll}
		\hline
		Name & Duration     & Duration    & Redshift\\
		     & (T$_{90}$,s) & (T$_{X}$,s) &  \\
		\hline
		GRB 060111A & 13.2 & 3243  & 5.5\\
		GRB 060218A & 2100 & 2917  & 0.03\\
		GRB 121211A & 182  & 1415  & 1.023\\
		GRB 141031A	&	920  & 1100  &	--\\
		GRB 141121A & 1410 & < 5000\footnote{The start of the fast X-ray decay is missing in the XRT data due to a gap in the observation. We can set only an upper limit.} & 1.47\\
		GRB 140413A & 140  & 3899  & --\\
		GRB 161129A & 35.5 & 2000  & 0.645\\
		\hline
	\end{tabular}
	\end{table}

\begin{table}
	\centering
	\caption{The control sample of long GRBs selected with $10^3$s $>$ T$_x > 500$s. We list selected properties.}
	\label{table_control}
	\begin{tabular}{clll}
		\hline
		Name & Duration     & Duration    & Redshift\\
		     & (T$_{90}$,s) & (T$_{X}$,s) &  \\
		\hline
GRB 081028A & 260 & 529 &	3.038\\
GRB 090417B & 260 & 535 &	0.35\\
GRB 111016A	&	550 & 900 &	6.4\\
GRB 111123A & 290 & 647 &	3.15\\
GRB 111215A	&	796 & 990 &	2.06\\
GRB 121217A	&	778 & 720 &	3.1\\
GRB 130606A & 276 & 729 &	5.91\\
GRB 140114A & 140 & 578 &	3.0\\
GRB 150616A	&	600 & 618 & --\\
		\hline
	\end{tabular}
\end{table}

It can already be observed that our gold bursts are all located at small redshifts, when compared to the other samples. \citet{gen13} discussed this fact, and explained it as a selection effect due to the Swift trigger conditions: for more distant events the satellite would rather trigger only on the peak of the emission, thus reducing the recorded T$_{90}$. In addition, other classes of GRBs are also a very low redshift while not being ultra-long ones \citep[e.g.][]{der17}. 

One may argue that due to the time dilation, the control sample may be biased toward distant events. From Table \ref{table_control}, we can see that the redshifts of those bursts range from 0.35 to 6.4, and that most of the events are located close to the mean redshift value of 2.8 reported by \citet{jak06} for normal long GRBs. We thus assume that sample not to be biased against high distance. Also, as reported during the introduction, we are looking for a way to classify ulGRBs within minutes, while the redshift measurement requires hours: we are not supposed to have access to this information for our study.

\section{Spectral properties}
\label{spectral}

\subsection{Prompt spectrum}
We retrieved from the {\it Swift} BAT archive all the prompt data related to our three samples, and performed a spectral fitting. In order to have a coherent sample, each data set has been reprocessed by the task {\it batbinevent} to create the spectrum. We used the version 6.22 of the FTOOLS with the latest version of the Swift calibration database. As we were looking for differences in the first minutes of the events, their spectrum during the first 300 seconds of the prompt phase was extracted. For normal long events, we also extracted the whole spectrum to decipher if differences between the first part of the event and the whole burst existed. Note that several events are lasting in the BAT band less than 300 seconds, while still fulfilling our filtering criteria in the XRT band.

As one can expect due to the narrow spectral band of the BAT instrument, complex models such as the Band model \citep{ban93} do not provide well constrained fits. So, we concentrated on an estimation of the hardness, using a single power law model. The results are listed in Table \ref{table_spectral}. The poor fit of GRB 170714A seems to be due to a larger background as compared to the other events. Indeed, a visual inspection of the residuals showed some random variability in the spectrum. All the other bursts are well fit by the single power law model.

\begin{figure}
	\centering
		\includegraphics[width=9cm]{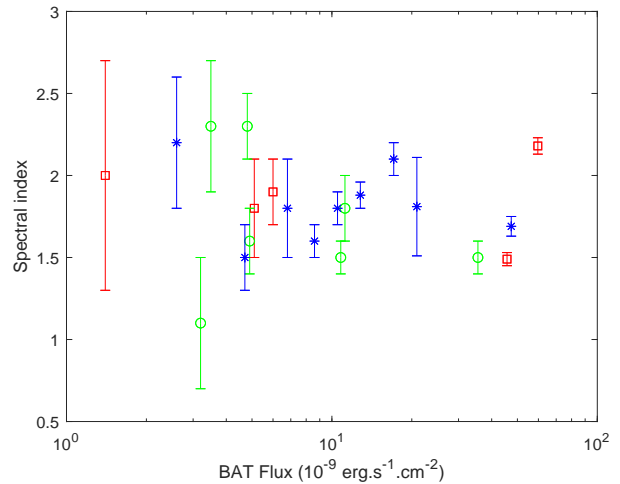}
	\caption{Spectral index as a function of the initial flux of the burst. We present the gold sample in red, the silver sample in black, and the control sample in blue. The flux has been extracted for the first 300 seconds of the burst only, and is expressed in the 15-150.0 keV band.\label{fig_alpha}}
\end{figure}

We plot in Fig. \ref{fig_alpha} the spectral index as a function of the mean flux during the first 300 seconds. One can clearly see a standard effect of the brightness, i.e. that faint prompt phases have a low flux and large error bars. A statistical test that accounts for the size of the error bars and the small number of events indicated that the gold and control samples are statistically similar. We did not find any discrepancies within the flux distributions of the three samples.

\begin{table*}
	\centering
	\caption{Spectral properties of our various samples. For the gold and silver samples, we extracted the spectra during the first 300s. For the control sample, we extracted two spectra when possible: one during the first 300s, and a second one for the whole duration of the event. We report both results with the integration time in the Table.}
	\label{table_spectral}
	\begin{tabular}{cccccccc}
		\hline
   burst    & Extraction start & Integration & Spectral & flux                                & Reduced  & d.o.f. & sample \\
		        & UTC time (s)     &  time (s)   & index    & (10$^{-9}$ erg. cm$^{-2}$.s$^{-1}$) & $\chi^2$ &        &        \\
\hline
GRB 101225A & 18:34:53 & 300 & 2.0  $\pm$ 0.7  & 1.4  $\pm$ 0.4 & 0.80 & 59 & Gold \\
GRB 111209A & 07:12:16 & 300 & 1.49 $\pm$ 0.04 & 45.7 $\pm$ 2.9 & 0.84 & 59 & Gold \\
GRB 121027A & 07:32:40 & 300 & 1.9  $\pm$ 0.2  & 6.0  $\pm$ 0.5 & 0.74 & 59 & Gold \\
GRB 130925A & 04:11:36 & 300 & 2.18 $\pm$ 0.05 & 59.7 $\pm$ 5.2 & 0.71 & 59 & Gold \\
GRB 170714A & 12:25:52 & 300 & 1.8  $\pm$ 0.3  & 5 $^{+9}_{-3}$ & 1.36 & 56 & Gold \\
\hline
GRB 060111A & 04:23:07 & 300 & 1.6  $\pm$ 0.2  & 4.9  $\pm$ 0.4 & 0.66 & 59 & Silver \\
GRB 060218A & 03:34:32 & 300 & 2.2  $\pm$ 0.2  & 4.8  $\pm$ 0.5 & 0.91 & 59 & Silver \\
GRB 121211A & 13:47:13 & 300 & 2.4  $\pm$ 0.4  & 3.4  $\pm$ 0.5 & 0.96 & 59 & Silver \\
GRB 140413A & 00:09:52 & 300 & 1.5  $\pm$ 0.1  & 35.5 $\pm$ 3.0 & 0.74 & 56 & Silver \\
GRB 141031A & 07:18:39 & 300 & 1.2  $\pm$ 0.4  & 3.2  $\pm$ 0.6 & 0.80 & 59 & Silver \\
GRB 141121A & 03:50:56 & 300 & 1.8  $\pm$ 0.2  & 11.2 $\pm$ 1.0 & 0.99 & 59 & Silver \\
GRB 161129A & 07:11:57 & 300 & 1.5  $\pm$ 0.1  & 10.8 $\pm$ 0.4 & 1.53 & 56 & Silver \\
\hline
GRB 081028A & 00:25:04 & 260 & 1.87 $\pm$ 0.08 & 12.8 $\pm$ 0.9 & 0.86 & 59 & Control \\
GRB 090417B & 15:20:08 & 260 & 2.2  $\pm$ 0.4  & 2.6  $\pm$ 0.4 & 0.99 & 59 & Control \\
GRB 111016A & 18:37:12 & 300 & 1.8  $\pm$ 0.2  & 10.5 $\pm$ 0.8 & 0.72 & 59 & Control \\
            &          & 550 & 1.9  $\pm$ 0.1  & 6.4  $\pm$ 0.5 & 0.78 & 59 & Control \\
GRB 111123A & 18:13:29 & 300 & 1.69 $\pm$ 0.06 & 20.9 $\pm$ 1.5 & 0.52 & 59 & Control \\
            &          & 290 & 1.68 $\pm$ 0.06 & 21.5 $\pm$ 1.6 & 0.51 & 59 & Control \\
GRB 111215A & 14:04:16 & 300 & 1.8  $\pm$ 0.3  & 6.7  $\pm$ 0.9 & 0.88 & 59 & Control \\
            &          & 796 & 1.6  $\pm$ 0.2  & 4.1  $\pm$ 0.5 & 1.30 & 59 & Control \\
GRB 121217A & 07:17:58 & 300 & 1.5  $\pm$ 0.2  & 4.7  $\pm$ 0.4 & 0.97 & 59 & Control \\
            &          & 778 & 1.61 $\pm$ 0.09 & 7.0  $\pm$ 0.5 & 0.79 & 59 & Control \\
GRB 130606A & 21:04:50 & 277 & 1.6  $\pm$ 0.1  & 8.6  $\pm$ 0.7 & 1.16 & 59 & Control \\
GRB 140114A & 11:57:52 & 140 & 2.1  $\pm$ 0.1  & 17.1 $\pm$ 1.2 & 0.90 & 59 & Control \\
GRB 150616A & 22:49:33 & 300 & 1.69 $\pm$ 0.06 & 47.4 $\pm$ 2.7 & 0.78 & 59 & Control \\
            &          & 600 & 1.72 $\pm$ 0.06 & 27.1 $\pm$ 1.6 & 1.08 & 59 & Control \\
		\hline
	\end{tabular}
\end{table*}

\subsection{Integrated energy}

\begin{figure}
	\centering
		\includegraphics[width=9cm]{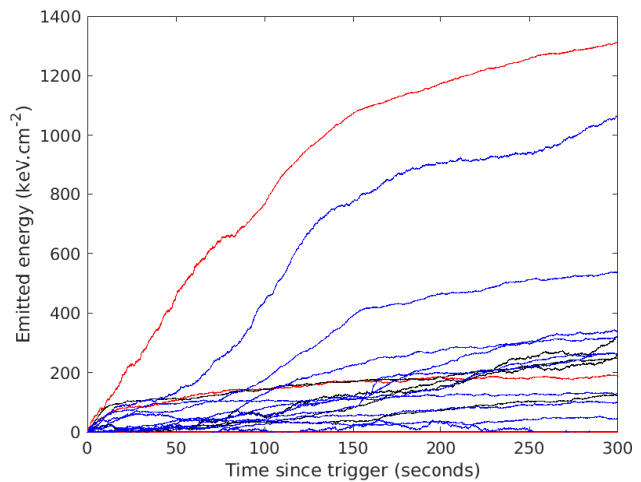}
	\caption{Fluence of the bursts as a function of time. The colors are the same than in Fig. \ref{fig_alpha}. The fluence has been extracted for the first 300 seconds of the burst only, and is expressed in the 15-150.0 keV band. Note that we are using the Swift trigger time when available, and thus some ultra-long events are quiescent during the temporal window used.\label{fig_fluence}}
\end{figure}

Ultra-long GRBs have larger energetic budgets, so one could ask if such is built differently than for normal long events? In Fig. \ref{fig_fluence}, we present the fluence of our bursts versus time. As can be seen therein, no clear pattern emerges. Ultra-long GRBs (gold sample) have similar energy emission rates as compared to normal long GRBs. One can note that one of the burst fluence plotted in this figure seems to decrease, which is not physical. This effect is due to the fact that we have used for our study only the information available {\it at the moment of the trigger}. For that event, the final background signal was found to be lower than the initial one: this non physical behavior is only due to a poor background correction.

The first derivative of the fluence (i.e. its growth rate) as a function of time provides a similar view as above, with no clear distinction between between long and ultra-long bursts. We, thus, conclude that the energy emission method is the same in these events.

\section{Temporal properties}
\label{timing}

Lastly, the light curves of ulGRBs were investigated for patterns atypical to those of normal long GRBs. We found no evidence to support such, using either an FFT transformation nor direct visual inspection. In the standard model view, this is perfectly plausible, as the prompt phase temporal profile is due to internal shocks not linked to the nature of the progenitor.

We also considered the distribution of burst durations. As already noted, ulGRBs have by definition durations of at least 1000 seconds, and can be classified with surety once 5000 seconds have elapsed without seeing a steep decline in their X-ray light curves. It is thereby possible to compute a probability that a burst is an ultra-long event using the information that a burst is still in its prompt phase. Indeed, the probability that a burst lasts more than X seconds knowing that it is active since Y seconds increases with the value of Y. We set X to be 5000 seconds, and tested various values of Y using the duration ($T_x$) distribution. These probabilities are listed in Table \ref{table_duration}.

As observed in Table \ref{table_duration}, if a burst lasts at least $\sim 20$ minutes (1200s), then the probability that it is an ulGRB is 50\%. For consistency, we included the case X = Y = 5000 seconds, which obviously produces a probability of 1.0. We also note that these probabilities are likely underestimated, because X should be fixed by the number of ulGRBs in our silver sample and not arbitrarily. However, as no unique indicator was found that could sufficiently classify ulGRBs within our silver sample, it is impossible to better constrain the value of X.

\begin{table}
	\centering
	\caption{Probability that a burst is an ulGRB as a function of the minimal observed emission duration. If the emission of the burst last at least this value, then the burst could be classified as an ulGRB (assuming the emission will continue) with the associated probability.}
	\label{table_duration}
	\begin{tabular}{cc|cc}
		\hline
		Duration    & Probability & Duration    & Probability\\
		(T$_{X}$,s) &             & (T$_{X}$,s) &  \\
		\hline
		200         &  0.0633     & 1200        & 0.4545 \\
		400         &  0.1389     & 2400        & 0.5556 \\
		600         &  0.2778     & 3600        & 0.8333 \\
		1000        &  0.4167     & 5000        & 1.0    \\
		\hline
	\end{tabular}
\end{table}

\section{Discussion \& Conclusions}
\label{discu}

\subsection{Can we define a burst to be ultra-long without the knowledge of the duration?}

\citet{boe15} already demonstrated that long and ultra-long GRBs were different, based on their duration distribution. The boundary between these two classes is however still not sure. \citet{gen13} proposed to use $10^4$ seconds as a cut-off limit, while \citet{boe15} suggested $10^3$ seconds. We defined this limit as $5\times 10^3$ seconds given that it resulted in a pure sample of ulGRBs, and because a significantly large sample of normal long GRBs with the presence of a few miss-identified ulGRBs would have no consequences on statistical studies.

From the first few minutes of the long and ultra-long GRBs studied here, none of the spectral or temporal properties tested were conclusive for discriminating these two burst classes. In fact, as it is clearly seen in Fig \ref{fig_alpha}, it is impossible to assess if the bursts in our silver sample are either ultra-long or normal long events. Thereby, it seems that there is no {\it a priori} parameter allowing the classification of ulGRBs before the end of their prompt phase.

{\it Swift}-BAT is clearly a limited instrument for a study such as ours, because its response bandwidth in the gamma-ray band is not very wide. We recognize that it is plausible that the/a ``correct" measurement for early phase ulGRB classification may have been overlooked by this work. For instance, \citet{pir14} found a blackbody component in the X-ray spectrum of GRB 130925A, and a re-analysis of GRB 111219A by \citet{gen13} revealed an extra component in its XMM-Newton spectra, initially classified as non-thermal, that was in fact more plausibly a thermal component. Such components were also found in the observations of GRB 170714A (Piro et al. in preparation). However, this component was too faint to be detected by {\it Swift}-XRT, and we are limited to the observable data tested in this paper.

Clearly, even a couple of prompt phase ulGRB observations from large facilities would significantly help in understanding if an empirical indicator exists for their early classification. The probabilities listed in Table \ref{table_duration} show that if an event is still active after $\sim 20$ minutes, there is $50\%$ of chance it will still be active for at least 2-3 hours. Such information can be easily obtained by the {\it Swift} team and disseminated as an automated GCN, flagging them as potential ultra-long events (it should be noted that any event with an activity of $\sim 20$ minutes is already inside our silver sample), to allow for a deep and fast follow-up.

\subsection{What are the consequences for the standard model?}
Our findings are important for the study of the central engine and progenitor of ulGRBs. \citet{iok16} have indicated that none of the models can be totally ruled out by the current observations. So it is possible that the progenitors of long and ultra-long GRBs are intrinsically different. However, both in the prompt phase (this work) and in the afterglow phase \citep{str13}, long and ultra-long events are extremely similar. Other studies \citep[e.g.][]{att17}, focusing on standard properties have also found that normal long GRBs and ulGRBs were behaving similarly. This clearly indicates that the central engines, albeit being active longer for ulGRBs, have the same properties in both cases and may be similar.

A consequence of similar long and ultra-long GRB central engines are obvious: the various progenitor models have to produce the same "class" of central engines. If one considers the magnetar model \citep{uso92} for the central engine of long GRBs, then obviously it is also favored for ulGRBs. However, in such a case, the criticisms formulated in \citet{iok16} and \citet{gen13} against this model (mostly the energy budget) still hold.

The white dwarf tidal disruption events defended by \citet{iok16} may also be plausible in the case of very close encounters, albeit their model uses a black hole of $10^5$ M$_\odot$. However, one would then need to explain how to produce normal long GRBs with such a
central engine, compatible with the various and numerous supernovae associations reported in the literature \citep[e.g.][]{kaw03,sta03}. Because this progenitor is a known emitter of gravitational waves in the band of {\it LISA} \citep{ann18}, GW observations could play a key role to validate this hypothesis.

Lastly, if one considers the collapsar model \citep{woo93} for long GRBs, we are dealing with a stellar mass black hole as the central engine, accreting the remains of the star. Our findings would favor this hypothesis. In particular, given that the difference in duration would only be due to the difference in size of the stars, as already argued in \citet{gen13}.

\subsection{Possible bias on our analysis}

As indicated in the introduction, several models could explain ulGRBs. For instance, if long GRBs are a result of the collapsar model of \citet{woo93}, while ulGRBs are due the white dwarf tidal disruption model of \citet{mac14}, a difference in the temporal and spectral regimes of long GRBs versus ulGRBs would be expected. That is, the collaspar model is based on chaotic emission of matter inside a jet, and emission from a white dwarf disruption is linked to an accretion disk. As we found no evidence of different observational characteristics to suggest the same global model, speculations on various models that could explain either of these types of GRBs were deferred to this section. 

As we stated earlier, long GRB and ulGRB populations reside at different distances from Earth. Thus, one could question if redshift effects mask observational differences between these two populations. The effects of redshifts on observations relate to time dilation, peak energy position, and the flux observed in a fixed band in the observer frame. The first two of these imply a factor of $(1+z)^{-1}$ in the temporal and $(1+z)$ in the spectral range. If we consider mean values, reshift effects indicate a factor $\Delta = (1+1)/(1+3) = 0.5$ exists between two populations. In the temporal domain, no power diagram shifts were found that would be interpreted as evidence to favor such effects. On spectral aspects, redshift effects should divide in half the peak energy value ($E_p$), and  lead to the appearance of a harder spectral slope of ulGRBs relative to long GRBs, if all the intrinsic values of $E_p$ for long GRBs were in the $80-100$ keV range.  We rule out this hypothesis by pointing out that the statistical studies of Fermi \citep{yu16}, or \citet{ama06} showed that $E_p$ peaks at larger values with a very broad distribution. 

The redshift effect on observed fluxes would be apparent in Fig. \ref{fig_fluence}, as a decreasing growth rate of the blue curves, compared to the red ones. It is beyond this study to suggest an estimate of how this effect would change the results of this figure. We speculate that a study based on the rest-frame properties, however, could shed more light on the similar and/or differing central engine properties of long GRBs and ulGRBs. From this work, a naive estimate based solely on the distance would be inaccurate as the {\it detection bandwidth} is also affected by redshifts, and the k-correction needed for clarifying it is highly model dependent. In addition, obviously the selection biases of ulGRB selection we discussed in the introduction would also require correction, as there would be a whole fraction of distant ulGRBs not present on Fig. \ref{fig_fluence} then. Particularly, if one considers that Pop III stars would lead to ulGRBs \citep{suw11, nag12, mac13}. Again, these would be hard to locate on our Fig. \ref{fig_fluence}, as we don't precisely know their properties: the fact they were not detected may not {\it only} be related to their brightness but {\it also} the triggering algorithm of the detecting instrument being less sensitive to their spectral or temporal properties \citep{gen13}, or to a redshift so large that the BAT band would be located above a few MeV in the rest frame. The only conclusion we can reach here, is that some events have similar redshifts (GRB 130925A and GRB 090417B, for instance), and similar properties in each of our tests, while belonging to two diverse groups. As no correction could change this result, we are confident that our work, even if affected by in redshift differences, provides a good representation of the true initial distributions.

\subsection{The future of the science of ulGRBs}

In this paper, we concentrated on possible markers available near the {\it Swift} trigger time to classify GRBs as ultra-long types.
Other tests are possibles, and could lead to new questions. For instance, a simple comparison of the BAT duration and the $T_X$ measurement (similar to a duration in the X-ray band) clearly indicates that ulGRBs should be soft events, and one could ask if these events are not in fact ulXRFs. However, the access of the hardness ratio is currently only available a few hours after the trigger. {\it Swift} is thus not the best suited observatory to answer these kinds of questions. However, in the future we may have more success in addressing such with new instruments in preparation.

{\it SVOM} \citep{gon18, wei16}, to be launched in a couple of years, is well suited for ulGRB studies. Most of all because it will observe the same direction for long periods (typically several hours), and possesses an image trigger extending to 20 minutes and possibly longer that will simplify efforts to detect and recognize ulGRBs \citep{dag18}. The multiwavelength capability of {\it SVOM} will allow the prompt emission to be monitored simultaneously in the visible (GWAC), hard X-rays (ECLAIRs), and gamma-rays (GRM), while for the afterglow emission this will be in NIR and visible (GFTs and VT), and in X-rays (MXT)---providing detailed diagnostics to the ulGRB class of events.

In the more distant future, THESEUS \citep{ama18} will provide sky surveys over a very wide energy band (0.3 keV $-$ 20 MeV) for the detection of ulGRBs with an unprecedented large sensitivity. Such is very important for faint ulGRBs, as it will provide very good resolution of their recorded spectra and light curves \citep{str18}. The {\it SVOM} and THESEUS experiments may allow us to carry out tests that we were unable to perform in this study. For instance, the monitoring of the soft to hard X-ray hardness evolution during the prompt emission (i.e. since the beginning of the burst). 

Lastly, as indicated previously, several proposed ulGRB progenitors are also gravitational wave (GW) emitters. Thereby, the next generation of GW instruments (the Einstein Telescope and possibly LISA), which will have an horizon encompassing the mean distance of ulGRBs \citep{pun10}, will provide unprecedented multi-messenger studies. We cannot exclude the possibility that the markers we are looking for appear more clearly in the gravitational wave signal.

\section*{Acknowledgements}
We would like to thank the anonymous referee who helped improving the text of this paper. We gratefully acknowledge support through NASA-EPSCoR grant NNX13AD28A. B.G. and N.B.O. also acknowledge financial support from NASA-MIRO grant NNX15AP95A, and NASA-RID grant NNX16AL44A. Parts of this research were conducted by the Australian Research Council Centre of Excellence for Gravitational Wave Discovery (OzGrav), through project number CE170100004. This research has been partly made under the auspices of the FIGARONet collaborative network supported by the Agence Nationale de la Recherche, program ANR-14-CE33. We thank Eric Howell and David Coward for useful discussion during the preparation of this article.





\begin{thebibliography}{99}
\bibitem[Anninos et al.(2018)]{ann18} Anninos, P., Fragile,  P. C., Olivier, S. S., Hoffman, R., Mishra, B., \& Camarda, K., \apj 865, 3, 2018
\bibitem[Amati(2006)]{ama06} Amati, L., MNRAS, 372, 233, 2006
\bibitem[Amati et al.(2018)]{ama18} Amati, L., O'Brien, P., G\"otz, D., et al., AdSpR 62, 191, 2018
\bibitem[atteia et al.(2017)]{att17} Atteia, J. -L., Heussaff, V., Dezalay, J. -P., et al., \apj, 837, 119, 2017
\bibitem[Band et al.(1993)]{ban93} Band, D., Matteson, J., Ford, L., et al., \apj 413, 281, 1993
\bibitem[Barat et al.(1992)]{bar92} Barat, C., Dezalay, J.P., Talon, R., et al., AIPC, 265, 304, 1992
\bibitem[Bo\"er et al.(2015)]{boe15} Bo\"er, M., Gendre, B., \& Stratta, G., \apj, 800, 16, 2015
\bibitem[Burrows et al.(2011)]{bur11} Burrows, D.N., Kennea, J.A., Ghisellini, G., et al., Nature, 476, 421, 2011
\bibitem[Cucchiara et al.(2015)]{cuc15} Cucchiara, A., Veres, P., Corsi, A., et al., \apj 812, 122, 2015
\bibitem[Dagoneau et al.(2018)]{dag18} Dagoneau, N., Schanne, S., Gros, A., \& Cordier, B., SF2A-2018: Proceedings of the Annual meeting of the French Society of Astronomy and Astrophysics, 2018
\bibitem[Dereli et al.(2017)]{der17} Dereli, H., Bo\"er, M., Gendre, B., Amati, L., Dichiara, S., \& Orange, N.B., \apj 850, 117, 2017
\bibitem[Eichler et al.(1989)]{eic89} Eichler D. et al., 1989, Nat, 340, 126
\bibitem[Fishman et al.(1994)]{fis94} Fishman, G.J., Meegan, C.A., Wilson, R.B., et al., ApJS 92, 229, 1994
\bibitem[Gehrels et al.(2004)]{geh04} Gehrels, N., Chincarini, G., Giommi, P., et al., \apj 611, 1005, 2004
\bibitem[Gendre et al.(2013)]{gen13} Gendre, B., Stratta, G., Atteia, J.L., et al., \apj, 766, 30, 2013
\bibitem[Gonzalez \& Yu(2018)]{gon18} Gonzalez, F. \& Yu, S., SPIE 10699E, 20, 2018
\bibitem[Greiner et al.(2015)]{gre15} Greiner, J., Mazzali, P.A., Kann, D.A., et al., Nature, 523, 189, 2015
\bibitem[Ioka et al.(2016)]{iok16} Ioka, K., Hotokezaka, K., \& Piran, T., \apj 833, 110, 2016
\bibitem[Jakobsson et al.(2006)]{jak06} Jakobsson, P., Levan, A., Fynbo, J.~P.~U., et al., \aap, 447, 897, 2006
\bibitem[Kawabata et al.(2003)]{kaw03} Kawabata, K. S., Deng, J., Wang, L., et al. \apj, 593, L19, 2003
\bibitem[Klebesadel et al.(1973)]{kle73} Klebesadel, R., Strong, I., \& Olson, R., \apj 182, L85, 1973
\bibitem[Levan et al.(2014)]{lev14} Levan, A.J., Tanvir, N.R., Starling, R.L.C., et al., \apj 781, 13, 2014
\bibitem[Levan(2015)]{lev15} Levan, A.J., Journal of High Energy Astrophysics 7, 44, 2015
\bibitem[Lien et al.(2016)]{lie16} Lien, A., Sakamoto, T., Barthelmy, S. D., et al., \apj 829, 7, 2016
\bibitem[MacLeod et al.(2014)]{mac14} MacLeod, M., Goldstein, J., Ramirez-Ruiz, E., Guillochon, J., \& Samsing, J., \apj, 794, 9, 2014
\bibitem[MacPherson et al.(2013)]{mac13} Macpherson, D., Coward, D.M., \& Zadnik, M.G., \apj, 779, 73, 2013
\bibitem[Mazets et al.(1981)]{maz81} Mazets, E.~P., Golenetskii, S.~V., Ilinskii, V.~N., et al., \apss, 80, 3, 1981
\bibitem[Mazets et al.(1982)]{maz82} Mazets, E. P., Golenetskii, S. V., Gurian, Iu. A., \& Ilinskii, V. N., Ap\&SS, 84, 173, 1982
\bibitem[Meszaros(2006)]{mes06} Meszaros, P., RPPh 69, 2259, 2006
\bibitem[Nagakura et al.(2012)]{nag12} Nagakura, H., Suwa, Y., \& Ioka, K., \apj, 754, 85, 2012
\bibitem[Piro et al.(1998)]{pir98} Piro, L., Heise, J., Jager, R., et al., \aap 329, 906, 1998
\bibitem[Piro et al.(2014)]{pir14} Piro, L., Troja, E., Gendre, B., et al., \apj 790, L15, 2014
\bibitem[Punturo et al.(2010)]{pun10} Punturo, M., Abernathy, M., Acernese, F., et al., CQGra 27, 194002, 2010
\bibitem[Ricker et al.(2003)]{ric03} Ricker, G. R., Atteia, J.-L., Crew, G. B, et al., AIPC, 662, 3, 2003
\bibitem[Stanek et al.(2003)]{sta03} Stanek K.Z, Matheson, T., Garnavich, P.M., et al., \apj, 591, L17, 2003
\bibitem[Stratta et al.(2013)]{str13} Stratta, G., Gendre, B., Atteia, J.L., et al., \apj 779, 66, 2013
\bibitem[Stratta et al.(2018)]{str18} Stratta, G., Ciolfi, R., Amati, L., et al., AdSpR 62, 662, 2018
\bibitem[Suwa \& Ioka(2011)]{suw11} Y. Suwa, K. Ioka, \apj 726, 107, 2011
\bibitem[Usov(1992)]{uso92} Usov, V.V., Nature 389, 635, 1992
\bibitem[Wei et al.(2016)]{wei16} Wei, J., Cordier, B., Antier, S., et al., Proceedings of the Workshop held from 11th to 15th April 2016 at Les Houches School of Physics, France, 2016 [arXiv:1610.06892]
\bibitem[Woosley(1993)]{woo93} Woosley, S.E., \apj, 405, 273, 1993
\bibitem[Yu et al.(2016)]{yu16} Yu, H.-F., Preece, R.D., Greiner, J., et al., \aap, 588, 135, 2016
\bibitem[Zhang et al.(2014)]{zha14} Zhang, B.B., Zhang, B., Murase, K., Connaughton, V., \& Briggs, M.S.,\apj, 787, 66, 2014

\end{thebibliography}


\bsp	
\label{lastpage}
\end{document}